\documentclass[twocolumn,english,aps,prl,reprint,floatfix,notitlepage,footinbib,preprintnumbers]{revtex4-1}
\pdfoutput=1
\usepackage{lmodern}

\usepackage[T1]{fontenc}
\usepackage[latin9]{inputenc}
\usepackage{geometry}
\usepackage{color}
\usepackage{dsfont}
\usepackage{slashed}
\usepackage{ragged2e}
\usepackage{comment}
\usepackage{babel}
\usepackage{amsmath,amssymb,amsthm,enumitem}
\usepackage{graphicx}
\usepackage{stackengine}
\usepackage{esint}
\usepackage[unicode=true,pdfusetitle,
 bookmarks=true,bookmarksnumbered=false,bookmarksopen=false,
 breaklinks=false,pdfborder={0 0 1},backref=false,colorlinks=true]
 {hyperref}
\hypersetup{pdftitle={SMEFT_fermions},
 pdfauthor={Grant N. Remmen and Nicholas L. Rodd},
 citecolor=black,linkcolor=black,urlcolor=black}
\usepackage{breakurl}
\usepackage[hang,flushmargin]{footmisc} 
\allowdisplaybreaks
\makeatletter

\usepackage{etoolbox}
\apptocmd{\thebibliography}{\justifying\setlength{\leftskip}{7.4mm}}{}{}

 \@ifundefined{textcolor}{}
 {%
   \definecolor{BLACK}{gray}{0}
   \definecolor{WHITE}{gray}{1}
   \definecolor{RED}{rgb}{1,0,0}
   \definecolor{GREEN}{rgb}{0,1,0}
   \definecolor{BLUE}{rgb}{0,0,1}
   \definecolor{CYAN}{cmyk}{1,0,0,0}
   \definecolor{MAGENTA}{cmyk}{0,1,0,0}
   \definecolor{YELLOW}{cmyk}{0,0,1,0}
 }

\usepackage{babel}

\makeatletter
\def\simgt{\mathrel{\lower2.5pt\vbox{\lineskip=0pt\baselineskip=0pt
           \hbox{$>$}\hbox{$\sim$}}}}
\def\simlt{\mathrel{\lower2.5pt\vbox{\lineskip=0pt\baselineskip=0pt
           \hbox{$<$}\hbox{$\sim$}}}}
\makeatother

\newcommand{\ket}[1]{\left| #1 \right\rangle}
\newcommand{\bra}[1]{\left\langle #1 \right |}

\newcommand{\be}{\begin{equation}}
\newcommand{\ee}{\end{equation}}
\newcommand{\Refc}[1]{Ref.~\cite{#1}}
\newcommand{\Refcs}[1]{Refs.~\cite{#1}}
\newcommand{\Fig}[1]{Fig.~\ref{#1}}
\newcommand{\Eq}[1]{Eq.~(\ref{#1})}
      
\begin{document}

\pagestyle{plain}

\title{Flavor Constraints from Unitarity and Analyticity}

\author{Grant N. Remmen}
\author{Nicholas L. Rodd}
\affiliation{Center for Theoretical Physics and Department of Physics\\
University of California, Berkeley, CA 94720 and\\
Lawrence Berkeley National Laboratory, Berkeley, CA 94720}
\thanks{e-mail: \url{grant.remmen@berkeley.edu}, \\ \hspace{1cm} \url{nrodd@berkeley.edu}}


\begin{abstract}
\noindent 
We use unitarity and analyticity of scattering amplitudes to constrain fermionic operators in the standard model effective field theory.
For four-fermion operators at mass dimension 8, we scatter flavor superpositions in fixed standard model representations and find the Wilson coefficients to be constrained so that their contraction with any pair of pure density matrices is positive.
These constraints imply that flavor-violating couplings are upper-bounded by their flavor-conserving cousins.
For instance, LEP data already appears to preclude certain operators in upcoming $\mu \to 3e$ measurements.
\end{abstract}
\maketitle

\noindent{\it Introduction.}---Results from the Large Hadron Collider provide a remarkable affirmation of the Standard Model (SM).
Beyond the discovery of a SM-like Higgs, higher-order predictions have been validated to unprecedented levels, demonstrating that the SM provides an adequate description of many TeV-scale phenomena.
While these results bring into question naturalness arguments that had suggested new physics should emerge at these energies~\cite{Feng:2013pwa,Giudice:2017pzm}, they only strengthen the basic principles of quantum field theory, such as unitarity and analyticity, that underlie the SM.
In this Letter, we will demonstrate that no matter in what guise new physics ultimately appears, as long as it obeys these same basic tenets, there are nontrivial flavor constraints on the types of interactions it can produce.

If new physics is too heavy to be produced on-shell, it can still leave experimental imprints via off-shell states.
The appropriate language to describe observables in this scenario is an effective field theory (EFT) of the low-energy degrees of freedom, in this case of the SM.
At each mass dimension in the SMEFT, there is a basis of gauge and Lorentz invariant operators~\cite{Warsaw,Henning:2015alf}; new physics can be differentiated only through the specific values of the SMEFT Wilson coefficients, the last vestige of the ultraviolet (UV) completion.
The naive expectation may be that the SMEFT Wilson coefficients can take on any value consistent with current experimental constraints.
But this expectation is wrong.

Instead, infrared (IR) consistency---analytic and unitary properties of scattering amplitudes, causality, etc.---only holds for a subset of all possible EFT Lagrangians.
As shown in \Refcs{Nima,Pham:1985cr,Ananthanarayan:1994hf,Pennington:1994kc}, the $s^2$ coefficient of the forward scattering amplitude can, by virtue of analyticity and the optical theorem, be written as an integral over the cross section and hence must be positive, thereby constraining the Wilson coefficients.
This principle of bounding EFTs via IR consistency has been used to constrain a litany of theories, including fermionic scattering~\cite{Brando}.
However IR consistency bounds are only beginning to be systematically applied to the SMEFT; see \Refc{bosons} and references therein \footnote{Further discussion on positivity bounds can be found in \Refc{ZZ} and references therein, but see also the discussion on \Refc{ZZ} in \Refc{bosons}}.

In this Letter, we investigate the implications of these bedrock field theory principles for the fermionic sector of the SMEFT.
Since the optical theorem arguments of Refs.~\cite{Nima,Pham:1985cr,Ananthanarayan:1994hf,Pennington:1994kc} require amplitudes $\propto s^2$ in the forward limit, we must consider dimension-8 operators.
We demonstrate that for a large class of such operators, analyticity and unitarity place rigid constraints on the allowed flavor structure.
We consider operators of the schematic form $c_{mnpq} \partial^2 (\bar{\psi}_m \psi_n)  (\bar{\psi}_p \psi_q)$, where the indices $mnpq$ index flavor.
Consistency of the EFT will require that the Wilson coefficients $c_{mnpq}$ are positive when contracted with an arbitrary pair of pure density matrices.
The simplest consequence of this will be that flavor diagonal interactions must obey positivity, e.g., $c_{1111} > 0$.
Yet we also find that any flavor-violating interactions will be strictly bounded by a flavor-conserving analogue.
More generally, the full flavor space of operators will be subject to a complicated set of inequalities.
While the conventional challenge of detecting dimension-8 operators persists, these predictions allow for a test of whether an emerging new physics signal arises from a sector that is consistent with IR field theory axioms.
Detection of nonzero Wilson coefficients outside of the region allowed by our bounds would falsify analyticity (i.e., locality), unitarity, or Lorentz invariance in the UV.

We organize the remainder of this Letter as follows.
First we construct the basis of fermionic operators we will consider.
Next we compute scattering amplitudes for fermions in a superposition of generations and derive our family of positivity bounds, demonstrating the precise conditions imposed on the Wilson coefficients.
We explain the consequences of these constraints for flavor violation and demonstrate how these bounds are precisely satisfied in example UV completions.
Finally, we explore phenomenological implications.

\medskip

\noindent{\it Operators.}---We wish to consider dimension-8 four-fermi operators in the SMEFT~\footnote{We note that our approach is distinct from the progress made in establishing bounds on dimension-6 operators, for example see Refs.~\cite{Adams:2008hp,Low:2009di,Englert:2019zmt}, in that the latter results often invoke additional assumptions beyond unitarity and analyticity.}.
Our field content consists of the left-handed quark and lepton multiplets $Q$ and $L$ and the right-handed lepton $e$ and up- and down-type quark multiplets $u$ and $d$, where all quarks are triplets of ${\rm SU}(3)$ and $Q$ and $L$ are doublets of ${\rm SU}(2)$.
Each field carries a generation index running from $1$ to $N_f$, where in the SM $N_f = 3$. We implicitly sum over repeated flavor indices throughout.
For the purposes of the present Letter, we will restrict ourselves to consideration of scattering of eigenstates of the SM gauge group, which in the SMEFT means that we require the basis of operators containing at an even number of each type of fermionic field (modulo flavor).

Building a minimal basis of operators requires modding out by symmetries, spinor/tensor identities (Fierz, Schouten, Levi-Civita), completeness relations for ${\rm SU}(N)$ generators, integration by parts, and field redefinitions.
As in \Refc{bosons}, we will work in the unbroken phase of the SMEFT, meaning we can treat the fermions as effectively massless ($\slashed{\partial} \psi = 0$); this is equivalent to the assumption that the UV scale of the higher-dimension operators far exceeds SM fermion masses. 
As chirality and helicity coincide in the massless limit, we will scatter definite-helicity states of appropriate handedness.

Our basis of operators is given in Eqs.~(\ref{eq:OselfQ}--\ref{eq:OcrossQ2}). 
For fields $\psi_m$, let us define currents charged under the SM gauge group,
\be
\begin{aligned}
J^\mu[\psi]_{mn} &\!=\!{\bar\psi}_m \gamma_\mu \psi_n &\! J^{\mu}[\psi]^a_{mn} &\!=\! \bar\psi_m T^a \gamma_\mu \psi_n \\
 J^\mu[\psi]^I_{mn} &\!=\! \bar\psi_m \tau^I \gamma_\mu \psi_n &\!  J^{\mu}[\psi]^{Ia}_{mn} &\!=\! \bar\psi_m \!\tau^I \! T^a \gamma_\mu \psi_n,
\end{aligned}
\ee
where $\tau^I$ and $T^a$ are the generators of ${\rm SU}(2)$ and ${\rm SU}(3)$, respectively. The self-quartic, self-hermitian operators are then \footnote{Throughout this work, all operators should be understood as contributing to the effective Lagrangian.}:
\be 
\begin{aligned}\label{eq:OselfQ}
{\cal O}_1[\psi] & = -c_{mnpq}^{\psi ,1}\partial_{\mu} J_\nu[\psi]_{mn} \partial^\mu \! J^\nu[\psi]_{pq}, \!\!&\!\! & \psi=\text{any}\\
{\cal O}_{2}[\psi] & = -c_{mnpq}^{\psi,2}\partial_{\mu} J_\nu[\psi]^I_{mn} \partial^\mu \! J^{\nu}[\psi]^I_{pq}, \!\!&\!\! &\psi=L,\!Q\\
{\cal O}_{3}[\psi] & = -c_{mnpq}^{\psi,3}\partial_{\mu} J_\nu[\psi]^a_{mn} \partial^\mu \! J^{\nu}[\psi]^a_{pq},\!\! &\!\! &\psi=d,\!u,\!Q\\
{\cal O}_{4}[Q] & = -c_{mnpq}^{Q,4}\partial_{\mu} J_\nu[Q]^{Ia}_{mn} \partial^\mu \! J^{\nu}[Q]^{Ia}_{pq}.
\end{aligned}
\ee
Here, ``any'' denotes $\psi$ being able to take on each of the SM fermionic fields, $Q$, $L$, $e$, $u$, or $d$. The Wilson coefficients are written as tensors $c_{mnpq}$ in flavor space, taking complex values subject to the symmetrization condition $c_{mnpq}=c_{pqmn}$ and the self-hermitian condition $c_{mnpq}=c_{nmqp}^{*}$. Imposing both leaves $N_{f}^{2}(N_{f}^{2}+1)/2$ real operators for each choice of $\psi$ in each line.

The remaining operators we consider are self-hermitian cross-quartics and come in two types. First, 
\be 
\begin{aligned}\label{eq:OcrossQ}
\hspace{-0.2cm}{\cal O}_{J1}[\psi, \!\chi] &  \!=\!
-b_{mnpq}^{\psi\chi,1}\partial_{\mu} J_\nu[\psi]_{mq}\partial^{\mu}\! J^\nu [\chi]_{np}, \!\!&\!\! &\psi,\! \chi \! = \! \text{any}\\
\hspace{-0.2cm}{\cal O}_{J2}[Q,\! L] &\!=\!
-b_{mnpq}^{QL,2}\partial_{\mu}J_\nu[Q]^I_{mq}\partial^{\mu}\! J^\nu[L]^I_{np},\\
\hspace{-0.2cm}{\cal O}_{J3}[\psi,\! \chi] &\!=\!
-b_{mnpq}^{\psi\chi,3}\partial_{\mu} J_\nu[\psi]^a_{mq} \partial^{\mu}\! J^\nu[\chi]^a_{np}, \!\!&\!\! &\psi,\! \chi \! \in \! \{ d,\! u,\! Q\} ,\!\!\!\!\!\!\!\!\!\!\!\!\!\!\!\!\!\!
\end{aligned}
\ee
where in each line $\psi\neq \chi$  and $b_{mnpq}^{\psi\chi} = b_{nmqp}^{\chi\psi}$ so that ${\cal O}_J[\psi,\chi]={\cal O}_J[\chi,\psi]$.
Defining the singlet tensor $K_{\mu\nu}[\psi]_{mn} = \bar\psi_m \gamma_\mu D_\nu \psi_n$ and the fundamental tensors $K_{\mu\nu}[\psi]^I_{mn}= \bar\psi_m \tau^I \gamma_\mu D_\nu \psi_n$ and $K_{\mu\nu}[\psi]^a_{mn}= \bar\psi_m T^a \gamma_\mu D_\nu \psi_n$ \footnote{Note that $K$ is related to the Dirac stress-energy tensor: $T_{\mu\nu} \! \subset \! -i(K_{\mu\nu}[\psi]_{mm}\! + \! K_{\nu\mu}[\psi]_{mm}\! - \!g_{\mu\nu} K_\rho^{\;\;\rho}[\psi]_{mm})/2+{\rm h.c}$.}, we also have
\be 
\begin{aligned}\label{eq:OcrossQ2}
{\cal O}_{K1}[\psi, \!\chi] &  \!=\!
-a_{mnpq}^{\psi\chi,1}K_{\mu\nu}[\psi]_{mq}K^{\nu\mu}[\chi]_{np}, \!\!&\!\! &\psi,\! \chi \! = \! \text{any}\\
{\cal O}_{K2}[Q,\! L] &\!=\!
-a_{mnpq}^{QL,2}K_{\mu\nu}[Q]^I_{mq}K^{\nu\mu}[L]^I_{np}\\
{\cal O}_{K3}[\psi,\! \chi] &\!=\!
-a_{mnpq}^{\psi\chi,3}K_{\mu\nu}[\psi]^a_{mq} K^{\nu\mu}[\chi]^a_{np}, \!\!&\!\! &\psi,\! \chi \! \in \! \{ d,\! u,\! Q\},\!\!\!\!\!\!\!\!\!\!\!\!\!\!\!\!\!\!
\end{aligned}
\ee
again requiring $\psi\neq \chi$ in each line and $a^{\psi\chi}_{mnpq} = a^{\chi\psi}_{nmqp}$ so that ${\cal O}_K[\psi,\chi]={\cal O}_K[\chi,\psi]$. For each tensor $a$ and $b$, the entries are complex numbers, subject to the self-hermitian condition $a_{mnpq}=a_{qpnm}^{*}$. This gives $N_{f}^{4}$ real operators for each choice of $\psi,\chi$ on each line. For \Eq{eq:OcrossQ2}, in the cases where both fermion bilinears have the same chirality (e.g., ${\cal O}_{K1}[d,\! u]$), we can use Fierz identities to rewrite the operator into the $\partial J \partial J$ form at the cost of making SM gauge indices explicit:
\be
{\cal O}_{K1}[d,\! u] = -\frac{1}{2} a_{mnpq}^{du,1} \partial_\mu (\bar u_{n\sigma} \gamma_\nu d_{q\tau})\partial^\mu (\bar d_{m\tau} \gamma^\nu u_{p\sigma}),
\ee
writing $\sigma,\tau$ for fundamental ${\rm SU}(3)$ indices.
The counting of operators in Eqs.~(\ref{eq:OselfQ}--\ref{eq:OcrossQ2}) matches \Refc{Henning:2015alf} (which does not provide the explicit forms of the operators)~\footnote{Were we to consider scattering states that are superpositions of SM representations, we would need to also construct the basis of non-self-hemitian cross-quartics, which comprises both $B$-violating and -conserving operators, each containing an odd number of some of the fermionic fields ($e$, $d$, $u$, $Q$, $L$)}.

\begin{figure*}[t]
\begin{center}
\hspace{-0.38cm}
\includegraphics[width=2.09\columnwidth]{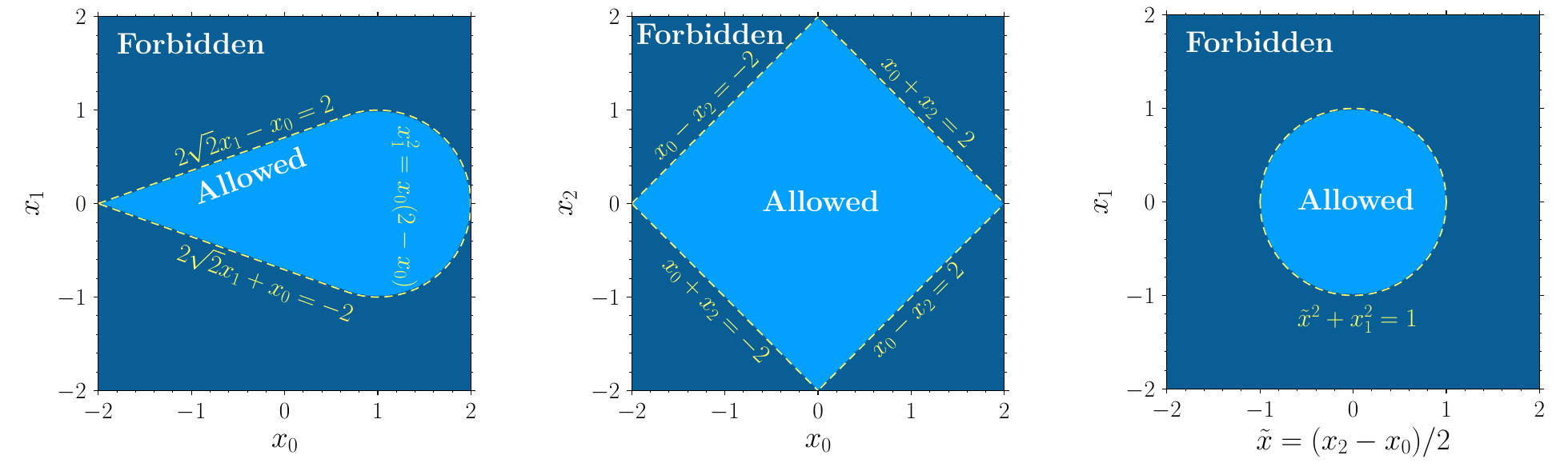}
\end{center}\vspace{-7mm}
\caption{Positivity bounds from \Eq{eq:cone} on the parameter space of a two-flavor theory.
In this example toy model, we enforce CP symmetry so that all $c^{e,1}_{mnpq} \in \mathbb{R}$.
For simplicity, in this illustration we further identify various Wilson coefficients, defining $c= c_{1111} = c_{2222} = c_{1221}$, $c_0 = c_{1122}$, $c_1 = c_{1112} = c_{1222}$, and $c_2 = c_{1212}$, so that $c,c_0$ are flavor-conserving, while $c_{1}$ ($c_2$) violates flavor by 1 (respectively, 2) units.
Consistency demands that $c>0$ and, defining $x_i=c_i/c$, that $-2+4|x_1|<x_0+x_2<2$ and $2|x_0-x_2|<2-x_0-x_2 + \sqrt{(x_0+x_2+2-4x_1)(x_0+x_2+2+4x_1)}$, for which projections are depicted above.
}
\label{fig:flavohedron}
\end{figure*}

\medskip

\noindent{\it Amplitudes and Bounds.}---We now wish to investigate the implications of imposing unitarity and analyticity of scattering amplitudes on the fermionic SMEFT operators discussed above. We first consider the  operators in \Eq{eq:OselfQ} and scatter states in an arbitrary superposition of flavors, but in afixed SM representation (i.e., a superposition of generations). Let us first scatter right-handed leptons via ${\cal O}_1[e]$; we take the states to be
\be
\begin{aligned}
\ket{\psi_1} &= \alpha_m \ket{\bar e_m} &\qquad \ket{\psi_2} &=  \beta_m \ket{e_m}\\
\ket{\psi_3} &= \gamma_m \ket{\bar e_m} &\qquad \ket{\psi_4} &= \delta_m \ket{e_m}\\
\end{aligned}\label{eq:state}
\ee
and require $\gamma_m = \beta_m^*$ and $\delta_m = \alpha_m^*$ for forward scattering (i.e., $\ket{\psi_1}\leftrightarrow \bra{\psi_4}$ and $\ket{\psi_2}\leftrightarrow \bra{\psi_3}$) \footnote{To define the overall sign of a fermionic amplitude, we also need to specify the order the external states. We choose $\ket{i} = \ket{1} \ket{2}$ and $\ket{f} = \ket{4} \ket{3}$ (recall we are working in the all incoming convention). This choice is made as in the elastic forward limit we then have $\ket{f} \to \ket{i}$ without any additional signs, allowing the optical theorem to be implemented straightforwardly.}.
Fixing helicities, we obtain the forward amplitudes~\footnote{Throughout this work, we only  state the $s^2$ contributions to the amplitude, as this is the relevant quantity we can bound. As in \Refc{bosons}, we consider a UV completion sufficiently weakly coupled that we can ignore contributions from diagrams with loops or multiple insertions of higher-dimension operators, which allows us to neglect contributions to \Eq{eq:amp} from operators of lower mass dimension; moreover, the SM contribution to this process will not diverge as $s^2$ by perturbative unitarity.}:
\be
{\cal A}(\bar e^-\! e^+ \! \bar e^- \! e^+)  \!=\! {\cal A} (\bar e^- \! \bar e^- \! e^+ \! e^+)\!=\! 4 c^{e,1}_{mnpq}\alpha_m \beta_n \beta_p^* \alpha_q^* s^2.\label{eq:amp}
\ee
Unitarity and analyticity then imply that 
$c^{e,1}_{mnpq}\alpha_m \beta_n \beta_p^* \alpha_q^* > 0$
for all vectors $\alpha$ and $\beta$.

Combining these vectors into matrices as $\rho^{\alpha}_{mq} = \alpha_m \alpha_q^*$ and $\rho^{\beta}_{np} = \beta_n \beta_p^*$, we define
\be
c_{\alpha \beta}^{e,1} = c^{e,1}_{mnpq} \rho^{\alpha}_{mq} \rho^{\beta}_{np}.
\ee
An analogous definition can be made for the remaining $a$, $b$, and $c$ tensors in Eqs.~(\ref{eq:OselfQ}--\ref{eq:OcrossQ2}). As $\rho^{\alpha}$ and $\rho^{\beta}$ are hermitian, idempotent, and of unit trace, they can be considered density matrices for pure states on a Hilbert space of dimension $N_f$, for which $\rho^{\alpha}_{mq} = \alpha_p \alpha_q^*$ represents the Schmidt decomposition. The $e^4$ bound can then be expressed as the requirement that $c_{\alpha \beta}^{e,1}$ is positive for every pair of pure density matrices $\rho^{\alpha}$ and $\rho^{\beta}$, i.e.,
\be
c^{e,1}_{\alpha \beta}> 0\;\;\; \forall\;\; \alpha,\beta.\label{eq:cone}
\ee
The space of Wilson coefficients satisfying this bound possesses nontrivial structure as illustrated in \Fig{fig:flavohedron}~\footnote{This criterion is reminiscent of the EFThedron bound on effective field theories~\cite{EFThedron}, which is related to the {\it spectrahedron} \cite{spectrahedron}, an object formed from a slice through the cone of positive {\it definite} matrices. Our requirement that the Wilson coefficients of the SMEFT be constrained in flavor space to be positive when contracted with any pair of pure density matrices could perhaps therefore be dubbed a {\it flavohedron}.}. 

For the $L^4$ operators in \Eq{eq:OselfQ}, we can write $\ket{\psi_1} = \alpha_{mi} L_{mi}$ and $\ket{\psi_2}=\beta_{mi}L_{mi}$, where the complex coefficients $\alpha$ and $\beta$ carry both a generation and fundamental ${\rm SU}(2)$ index. Using the generator completeness relation, the forward amplitudes ${\cal A}(\bar{\psi}^{+}\psi^{-}\bar{\psi}^{+}\psi^{-})$ and ${\cal A}(\bar{\psi}^{+}\bar{\psi}^{+}\psi^{-}\psi^{-})$ both equal
\be
\begin{aligned}
{\cal A} &=  4s^{2}\left[\left(c_{mnpq}^{L,1}-\tfrac{1}{4}c_{mnpq}^{L,2}\right)\alpha_{mi}^*\beta_{ni}\beta_{pj}^{*}\alpha_{qj}\right. \\& \hspace{22.6mm} \left.+\tfrac{1}{2}c_{mnpq}^{L,2}\alpha_{mi}^{*}\beta_{nj}\beta_{pj}^{*}\alpha_{qi}\right].
\end{aligned}
\ee
Marginalizing over all generation indices and ${\rm SU}(2)$ charges, we find that the bounds become
\be 
c_{\alpha\beta}^{L,1}+\tfrac{1}{4}c_{\alpha\beta}^{L,2}> 0\qquad\text{and}\qquad c_{\alpha\beta}^{L,2} > 0.\label{eq:coneL}
\ee
Proceeding analogously for $u^4$, $d^4$, and $Q^4$, we find:
\be
\begin{aligned}
&c_{\alpha\beta}^{u,1}+\tfrac{1}{3}c_{\alpha\beta}^{u,3}, &\;\;& c_{\alpha\beta}^{Q,1}+\tfrac{1}{4}c_{\alpha\beta}^{Q,2}+\tfrac{1}{3}c_{\alpha\beta}^{Q,3}+\tfrac{1}{12}c_{\alpha\beta}^{Q,4},\\
&c_{\alpha\beta}^{u,3}, && c_{\alpha\beta}^{Q,2}+\tfrac{1}{3}c_{\alpha\beta}^{Q,4}, \\
&c_{\alpha\beta}^{d,1}+\tfrac{1}{3}c_{\alpha\beta}^{d,3}, && c_{\alpha\beta}^{Q,3}+\tfrac{1}{4}c_{\alpha\beta}^{Q,4},\\
&c_{\alpha\beta}^{d,3},&& c_{\alpha\beta}^{Q,4}\\
&\text{are all}> 0.
\end{aligned}\label{eq:conec}
\ee

Let us now bound the cross-quartic operators. We start by scattering $d$ and $e$, so we take $\ket{\psi_{2,3}}$ as in \Eq{eq:state}, with $\gamma_m = \beta_m^*$, but take $\ket{\psi_{1,4}}$ to instead be $\alpha_{mi} \ket{\bar d_m}$ and $\alpha^*_{mi} \ket{d_m}$, respectively.
The forward amplitudes are
\be
{\cal A}(\bar e^{-} \! d^{+} \! \bar d^{-} \! e^{+}\! ) \! \!= \! {\cal A}(\bar e^- \! \bar d^- \! d^+ \! e^+\! )\!\! =\! a^{de,1}_{mnpq}\alpha_m \beta_{ni}^* \beta_{pi} \alpha_q^*  s^2 \! .
\ee
In analogy with \Eq{eq:cone}, we find that the combination $a_{\alpha\beta}^{de,1} = a^{de,1}_{mnpq} \rho^{\alpha}_{mq} \rho^{\beta}_{np}$ is positive:
\be
a^{de,1}_{\alpha \beta}  > 0.\label{eq:conea}
\ee
The operator ${\cal O}_{J1}[d,\! e]$ does not contribute to this amplitude in the forward limit. Analogously,
\be 
a^{ue,1}_{\alpha \beta} ,\, a^{eL,1}_{\alpha \beta} ,\, a^{dL,1}_{\alpha \beta} ,\, a^{uL,1}_{\alpha \beta} ,\, a^{eQ,1}_{\alpha \beta}  \,\text{are all}\,  > 0.\label{eq:conea2}
\ee
For the cross-quartic operators involving fermion bilinears with nontrivial ${\rm SU}(2)$ and ${\rm SU}(3)$ charges, we proceed as in the self-quartic case, marginalizing over the charges to find the necessary and sufficient bounds:
\be
\begin{aligned}
& a^{QL,1}_{\alpha \beta} \pm \tfrac{1}{4} a^{QL,2}_{\alpha \beta},  &\qquad & a^{du,1}_{\alpha \beta}  + \tfrac{1\pm 3}{12} a^{du,3}_{\alpha \beta} ,  \\
& a^{dQ,1}_{\alpha \beta}  + \tfrac{1\pm 3}{12} a^{dQ,3}_{\alpha \beta} , &  & a^{uQ,1}_{\alpha \beta}  + \tfrac{1\pm 3}{12} a^{uQ,3}_{\alpha \beta}, \\
& \text{are all}> 0.
\end{aligned}\label{eq:conea3}
\ee

\medskip

\noindent{\it Flavor Violation.}---The requirement in \Eq{eq:cone} has an important physical interpretation in terms of flavor. To demonstrate, let us again for simplicity consider $e^4$ operators (where for brevity we drop the $``e,1"$ superscript); our conclusions will apply to any of the positivity statements we prove in Eqs.~(\ref{eq:cone},$\,$\ref{eq:coneL},$\,$\ref{eq:conec},$\,$\ref{eq:conea}--\ref{eq:conea3}).
Our bounds imply that various flavor-conserving operators must have positive coefficient. For example, if $\alpha_m = \delta_{1m}$ and $\beta_m = \delta_{2m}$, we obtain $c_{1221}>0$, while if $\alpha_m = \beta_m = \delta_{1m}$, we find $c_{1111}>0$.
Moreover, if $\rho^{\alpha}$ or $\rho^{\beta}$ have off-diagonal support, we find components of $c_{mnpq}$ with magnitudes upper-bounded by their diagonal analogues. To illustrate, taking $\alpha_m = \delta_{1m}$ and $\beta_m = \delta_{2m} \cos \theta  + \delta_{3m} e^{i\phi} \sin\theta$ and marginalizing over $(\theta,\phi)$, we obtain:
\be
c_{1221} c_{1331} > |c_{1231}|^2.\label{eq:cone2}
\ee
This condition is notably similar to the completing-the-square condition we found in the context of CP violation in \Refc{bosons}; indeed, those bounds can be recast in the form of \Eq{eq:cone} by scattering superpositions of helicity (instead of flavor).
Here, however, the implication is that an interaction that violates lepton number such that $(\Delta L_\mu,\Delta L_\tau) = (+1,-1)$ is allowed, but only if the analogous flavor-conserving operators are also nonzero.
Similar statements hold for the relation between operators that violate and conserve other flavor quantum numbers, such as strong isospin or strangeness.
In fact, there is a connection to CP in our results.
Any CP-violating effects mediated by the operators considered here can be bounded as in \Eq{eq:cone2}, which can be seen as follows.
Violation of CP requires an imaginary coefficient for our operators, whereas the hermiticity condition demonstrates that this is only possible for flavor-violating couplings.
Accordingly, conservation and violation of both CP and flavor are connected, implying our results for flavor can be lifted to CP.
The result can be generalized further; it extends to the violation of any U(1) under which the fermions are charged.

The full set of relations between Wilson coefficients extends beyond \Eq{eq:cone2}, even for $N_f=2$ as shown in Fig.~\ref{fig:flavohedron}.
As depicted there, conditions also exist between certain flavor-conserving operators, such as $c > |c_0|/2$ for the example in the figure.
More generally, flavor conserving operators that induce scattering amplitudes $\propto st$ (which vanish in the forward limit), such as $c_{1122}$, are bounded by those that scale $\propto s^2$.

\medskip

\noindent{\it UV Completion.}---While a higher-spin coupling does not generically produce a UV completion with cutoff larger than the mass of the field, such a theory that does is that of a massive tensor with ghost-free Fierz-Pauli mass term coupled minimally to the energy-momentum, $\kappa \phi^{\mu\nu} T_{\mu\nu}$~\cite{Kurt,Cheung:2016yqr,Cheung:2019cwi}.
For simplicity, taking $T_{\mu\nu}$ to be that of the SM multiplet $e_m$, integrating out $\phi_{\mu\nu}$ generates ${\cal O}_1[e]$ with Wilson coefficients $c^{e,1}_{mnpq} = \kappa^2 (4\delta_{mq}\delta_{np} + \delta_{mn}\delta_{pq})/4m^2$.
As expected, virtual graviton exchange generates only flavor-conserving operators~\footnote{Note that flavor-violating interactions in theories of compact extra dimensions can be generated by the exchange of Kaluza-Klein vectors~\cite{Delgado:1999sv}.}.
Again our bounds are satisfied as $c^{e,1}_{\alpha \beta} = \kappa^2(4|\alpha|^2 |\beta|^2 + |\alpha\cdot \beta|^2)/4m^2$.
Such a Kaluza-Klein graviton is a generic feature of string compactifications and models of extra dimensions; see for example \Refc{PDG} and references therein.

\medskip

\noindent{\it Phenomenology.}---Projecting our bounds onto the space of experimental searches for beyond-the-SM (BSM) phenomena provides connections between naively disparate paths into the new physics landscape.
For example, nonzero $c_{1112}^{e,1}$ leading to the discovery of ${\rm Br}(\mu \to 3e)\sim 10^{-16}$--$10^{-12}$, as targeted by Mu3e~\cite{Blondel:2013ia,Crivellin:2016ebg}, would be in conflict with our bounds.
Writing $|c_{1112}^{e,1}| = \tilde{\Lambda}^{-4}$, a discovery would require $\tilde{\Lambda} \sim 100$--$300$ GeV.
Unitarity/analyticity imply $c^{e,1}_{1111} c^{e,1}_{2112} > |c^{e,1}_{1112}|^2$, so taking $c^{e,1}_{2112} \sim c^{e,1}_{1111} = \Lambda^{-4}$, there is a flavor conserving effect---potentially observable at a collider---that cannot be decoupled: $\tilde{\Lambda} > \Lambda$.
Collider measurements may be less sensitive than rare decay searches, but can be enhanced through the higher-dimension operator interference with the SM.
For fermion pair production at LEP, recast dimension-6 limits require $\Lambda \gtrsim 500$ GeV~\cite{Falkowski:2015krw,Alte:2018xgc}, in violation of our bounds.
These estimates are not a dedicated analysis, but highlight that collider constraints, together with our bounds, already impact flavor violation probes.

The above example used only the conical constraint of \Eq{eq:cone2}.
More generally, detection of nonzero SMEFT coefficients---and measurement of their signs via SM interference---would allow the richer structure in \Fig{fig:flavohedron} to be experimentally tested.
Wilson coefficients measured at different energies must be evolved to a common scale in order to apply analyticity bounds~\cite{Jenkins:2013zja,Jenkins:2013wua,Alonso:2013hga}, and a theory that satisfies the bounds at one scale should also do so deeper into the IR~\cite{Cheung:2014ega}.

While our bounds apply to dimension-8 operators, in many BSM scenarios dimension-6 four-fermi operators are induced as well, the latter giving rise to the leading deviations from the SM.
Dimension-8 effects are suppressed in the S-matrix by $(\Lambda_{\rm IR}/\Lambda_{\rm UV})^2$ with respect to dimension-6, where $\Lambda_{\rm UV}$ is the new physics scale and $\Lambda_{\rm IR}$ denotes the experimental probe.
Above, $\Lambda_{\rm IR} \sim m_{\mu}$ for $\mu \to 3 e$, whereas for $e^+ e^-$ collisions, larger values of $\Lambda_{\rm IR} \sim \sqrt{s}$ ($\sim$ electroweak scale for LEP) can be achieved by isolating the hard contribution to kinematic distributions.
Certain operators of dimension 6 and 8 can be disentangled by angular dependence in scattering~\cite{Alioli:2020kez} or other kinematic scaling differences, e.g., the high-$p_T$ tail of collider distributions (see, e.g., Refs.~\cite{Greljo:2017vvb,Fuentes-Martin:2020lea}).

Higher-scale probes are therefore particularly relevant for dimension-8 operators, even for rare flavor-changing processes where lower-energy measurements are the most sensitive at the level of the branching ratios.
For example, although ${\rm Br}(\tau \to 3e) \lesssim 10^{-8}$ is weaker than ${\rm Br}(\mu \to 3e) \lesssim 10^{-12}$, as $(m_{\tau}/m_{\mu})^4 \sim 10^5$ it probes a slightly higher scale.
Similarly, as $(m_t/m_{\mu})^4 \sim 10^{13}$, flavor-changing neutral current decays of the top are likely to be particularly sensitive probes~\cite{Fox:2007in,Sirunyan:2017uae,Aad:2019pxo}.
Beyond the dimension-8 qualifier, traditionally promising avenues for establishing BSM flavor violation such as neutral meson mixing remain sensitive~\cite{Isidori:2010kg,Isidori:2013ez}.
For flavor-conserving interactions, important probes are likely to be nonresonant dilepton or dijet events at colliders~\cite{Aaboud:2017yvp,Sirunyan:2017ygf,Aaboud:2017buh,Sirunyan:2018exx}, where the reach can be enhanced using events with $\Lambda_{\rm IR} \sim \sqrt{s}$~\cite{Domenech:2012ai,Alte:2017pme,Alte:2018xgc}.

This discussion is not exhaustive, but highlights how our bounds can give rise to striking correlations among experiments.
Higher-dimension operators among fermions have long been considered a compelling experimental target, including $T_{\mu\nu}T^{\mu\nu}$ interactions~\cite{PDG,Peskin} such as the Kaluza-Klein model we considered above, models of leptoquarks~\cite{Dorsner:2016wpm}, and fermion compositeness~\cite{Eichten:1983hw,Bellazzini:2017bkb,PDG2}.
Wilson coefficients of dimension-6 analogues of the operators in Eqs.~(\ref{eq:OselfQ}--\ref{eq:OcrossQ2}) have been bounded into the TeV scale via dijet, dilepton, diphoton, and top production measurements at the LHC~\cite{Aaboud:2017yvp,Sirunyan:2018wcm,Aaboud:2017yyg,Sirunyan:2018wnk,Sirunyan:2018ipj,Aad:2019fac,Zhang:2016omx}, as well as via neutrino scattering and parity violation~\cite{Falkowski:2017pss}.

Assumptions about the flavor structure of new physics, such as minimal flavor violation (MFV)~\cite{Chivukula:1987py,Hall:1990ac,DAmbrosio:2002vsn} or more general structures (e.g.,~\Refc{Bordone:2019uzc}), are independent of our bounds.
While MFV enforces certain flavor structures, it imposes no requirements on the signs of couplings, so MFV theories do not automatically obey our bounds.
Moreover, there are BSM flavor phenomena unprotected by our bounds, e.g., lepton universality; motivated by hints of lepton universality violation in measurements of $R_K$~\cite{Aaij:2014ora,Aaij:2019wad} and $R_{K^*}$~\cite{Aaij:2017vbb}, a scenario in which $a_{3112}^{QL,i} \neq a_{3222}^{QL,i}$ could generate unequal contributions to $\Gamma(B\!\rightarrow\!\! K^{(*)}\! e^+ e^-)$ and $\Gamma(B\!\rightarrow\!\! K^{(*)}\! \mu^+ \mu^-)$ while remaining consistent with our conditions. (For an analysis relevant to the SMEFT, see Ref.~\cite{Jenkins:2017jig}.)

This Letter leaves multiple avenues for future work.
While we have restricted our attention to scattering flavor superpositions in fixed SM representations, analyticity for superpositions of representations would lead to further connections between different operators, as in the case of bosons~\cite{bosons}.
This would allow for bounds on operators violating baryon and lepton number and connect the associated experimental searches.

\begin{acknowledgments}
We thank Cliff Cheung, Nathaniel Craig, Zoltan Ligeti, Aneesh Manohar, Yotam Soreq, Timothy Trott, and Mike Williams for feedback and discussions. G.N.R. and N.L.R. are supported by the Miller Institute for Basic Research in Science at the University of California, Berkeley.
The work of G.N.R. was performed in part at the Kavli Institute for Theoretical Physics, supported  by the National Science Foundation under Grant No. NSF PHY-1748958.
\end{acknowledgments}

\bibliographystyle{utphys-modified}
\bibliography{fermions}

\end{document}